\documentclass[useAMS,usenatbib]{mn2e}
\usepackage{amsmath}
\usepackage[dvips]{graphicx}

\usepackage{multirow}

\newcommand{\be}{\begin{equation}}
\newcommand{\ee}{\end{equation}}
\newcommand{\ba}{\begin{eqnarray}}
\newcommand{\ea}{\end{eqnarray}}
\newcommand{\no}{\nonumber \\}

\newcommand{\barr}{\begin{array}}
\newcommand{\earr}{\end{array}}

\newcommand{\solarmass}{\mbox{$\rm{M}_{\odot}$}}

\newcommand{\pdrv}[2]{ \frac{\partial #1}{\partial #2}}
\newcommand{\pdrvs}[2]{ \frac{\partial^2 #1}{\partial #2^2}}
\newcommand{\refeqn}[1]{Eq. \eqref{#1}}
\newcommand{\refeqns}[1]{Eqs. \eqref{#1}}

\title[Axially symmetric pseudo-Newtonian hydrodynamics code]
  {Axially symmetric pseudo-Newtonian hydrodynamics code}
\author[J. Kim et al.]
  {Jinho~Kim,$^1$\thanks{E-mail: jinho@astro.snu.ac.kr (JK); khi@astro.snu.ac.kr (HIK); choptuik@phas.ubc.ca (MWC); hmlee@snu.ac.kr (HML)}
  Hee Il Kim,$^{1,2}$\footnotemark[1] Matthew William Choptuik,$^{3,4}$\footnotemark[1] and Hyung Mok Lee$^1$\footnotemark[1] \\
  $^1$Department of Physics and Astronomy, FPRD, Seoul National University, Seoul, 151-742, Korea \\
  $^2$Korea Institute of Science and Technology Information, 245 Daehak-ro, Yuseong-gu, Daejeon, 305-806, Korea \\
  $^3$Department of Physics and Astronomy, University of British Columbia, Vancouver, Canada \\
  $^4$Canadian Institute for Advanced Research Cosmology \& Gravity Program}
\date{Released 2012 Xxxxx XX}

\pagerange{\pageref{firstpage}--\pageref{lastpage}} \pubyear{2012}

\begin{document}

\maketitle

\label{firstpage}

\begin{abstract}
We develop a numerical hydrodynamics code using a pseudo-Newtonian formulation that uses the weak field
approximation for the geometry, and a generalized source term for the Poisson equation that takes into account
relativistic effects. The code was designed to treat moderately relativistic
systems such as rapidly rotating neutron stars.
The hydrodynamic equations are solved using a finite volume method with High Resolution Shock Capturing
(HRSC) techniques.
We implement several different slope limiters for second order reconstruction schemes and also
investigate
higher order reconstructions such as PPM, ENO and WENO.
We use the method of lines (MoL) to convert the mixed spatial-time partial differential equations into
ordinary differential equations (ODEs) that depend only on time.
These ODEs are solved using second and third order Runge-Kutta methods.
The Poisson equation for the gravitational potential is solved with a multigrid method, and to
simplify the boundary condition, we use compactified coordinates which map spatial infinity to
a finite computational coordinate using a tangent function.
In order to confirm the validity of our code, we carry out four different tests including one and two
dimensional shock tube tests, stationary star tests of both non-rotating and rotating models and radial
oscillation mode tests for spherical stars.
In the shock tube tests, the code shows good agreement with analytic solutions which include shocks,
rarefaction waves and contact discontinuities.
The code is found to be stable and accurate: for example, when solving a stationary stellar model the
fractional changes in the maximum density,
total mass, and total angular momentum per dynamical time are found to be $3 \times 10^{-6}$, $5 \times
10^{-7}$ and $2 \times 10^{-6}$, respectively.
We also find that the frequencies of the radial modes obtained by the numerical simulation of
the steady state star agree very well with those obtained by linear analysis.
\end{abstract}


\section{Introduction}
It is necessary to take into account both special- and general relativistic effects in the studies of
the dynamics of compact astrophysical object such as neutron stars and black holes.
Some pulsars produce pulses of up to 1 KHz, corresponding to
rotation speeds at the surface of around $0.2c$.
Their typical sizes and masses are known to be around $10\rm{km}$ and $1.4\sim2\solarmass$, respectively,
giving compactness, $GM/Rc^2=0.2\sim0.3$. Therefore, a Newtonian approach cannot
properly describe neutron stars, even for the non-rotating case.

In general relativity, the dynamics of gravity (or spacetime) can be studied by solving the
 Einstein equations. The equations of motion for the matter are given, in part, by the conservation
law of the energy-momentum tensor which itself sources the gravitational field.
Computational approaches for solving general relativistic field equations constitute the
field of numerical relativity.

Over the  past few decades, many general relativistic hydrodynamic codes have been developed, starting
with \cite{wil72} who proposed a 3+1 Eulerian formulation (see also \citealp{wil03}).
Although Wilson's numerical approach was widely used to study problems such as
core collapse and accretion disks,
it produced large errors when fluid flows became ultra-relativistic  (\citealp{cen84};
\citealp{nor86}). In order to avoid these excessive errors, a new formulation was proposed by
\cite{mar91}.  This formulation makes it possible
to use existing numerical techniques
based on characteristic approaches for Newtonian hydrodynamics. In particular, these
include High Resolution Shock
Capturing (hereafter HRSC) methods that reduce the order of accuracy near shocks,
but minimize the amount of numerical dissipation. This dissipation is
very unnatural and can result in non-physical effects in
the numerical results.
Marti's formulation was extended to the general relativistic
case by the Valencia group (\citealp{fon00}), and this last work forms the basis for
most recent general relativistic hydrodynamical codes.
Recent reviews of the formulation and numerical methods can be
found in \cite{mar03} and \cite{fon08}.

However, when working in multiple spatial-dimensions, it still requires a lot of computational resources to
treat fluid dynamics in concert with the evolution of the general relativistic gravitational field.
In addition, numerical relativity simulations have frequently encountered instabilities which are
often associated with violations of the Hamiltonian and momentum constraints. (However, with the
development of new formulations which cast the Einstein equations in appropriate hyperbolic forms, as
well as the use of constraint-damping techniques,
significant progress has been made on this front: see~\citealp{sar12} for a very recent review
of this subject).
For these reasons, simulations using Newtonian gravity are still used even though they are not applicable to
very compact objects.

The aims of this paper are 1) to introduce a new formulation which applies a pseudo-Newtonian approach
(\citealp{kim09}) to the study  of moderately relativistic objects and 2) to
describe a numerical implementation of
this method. In our pseudo-Newtonian approach, which was introduced by \cite{kim09} for steady state models,
the gravitational field is treated by a weak field approximation, but special relativistic
effects are correctly taken into account. Specifically, the Newtonian gravitational potential that appears
in the weak field metric satisfies a Poisson equation, but the mass density that appears
as a source term for that equation is modified to include relativistic effects.
Of course this method cannot be applied to highly relativistic systems, but \cite{kim09} showed that
the pseudo-Newtonian formulation is valid for the modeling of mildly compact objects,
such as rotating neutron stars having surface rotation velocity up to
$\sim0.2c$ and compactness $\sim0.2$
(\citealp{kim09}). In this paper, we extend the pseudo-Newtonian approach to hydrodynamical
systems where the flows can be ultra-relativistic and gravity can be moderately strong.


The remainder of the paper is structured as follows:
In  section \ref{formulation}, we present the formulation and governing equations for
our system,  while the
numerical techniques employed in our study are given in section \ref{numer}.
We discuss various numerical tests of our code's treatment of hydrodynamics for the case of shock tubes in
 \ref{testshtb}, and for stationary stars in
\ref{teststatstar}.
A test which compares radial pulsation mode frequencies for polytropic stars determined
through dynamical evolution to those computed in linear theory is detailed in section \ref{testradmode}.
We conclude with a summary and discussion in section \ref{conclusion}.

Throughout this paper we use units in which $c=G=\solarmass=1$: these correspond a to unit time $=4.92\times
10^{-3}\rm{ms}$, unit length $=1.47\,\rm{km}$ and unit mass $=1.99\times 10^{33}\rm{g}$.

\section{Formulation}\label{formulation}
Our pseudo-Newtonian method was first discussed in the steady state context by \cite{kim09}. We assume
the weak field metric,
\be
\begin{split}
ds^2 &= g_{\mu\nu} dx^{\mu} dx^{\nu} \\
&= -(1+2\Phi)dt^2 + (1+2\Phi)^{-1}\delta_{ij}dx^{i}dx^{j},
\end{split}\label{eq1}
\ee
where $g_{\mu\nu}$ is the spacetime metric and $\Phi$ is the Newtonian gravitational potential.
With this metric, we neglect all higher order effects such as frame dragging and describe
gravity using  only a single
gravitational potential, just as in the Newtonian case.
The gravitational potential satisfies a Poisson equation with the active mass density,
$\rho_{\rm{active}}$, providing the source:
\be
\nabla^2 \Phi = 4\pi\rho_{\rm{active}}.\label{eq2}
\ee
The active mass density is computed from the relativistic definition of the energy.
For a perfect fluid, the energy momentum tensor can be expressed as
\be
T^{\mu\nu}=\rho_0 h u^{\mu}u^{\nu}+P g^{\mu\nu},\label{eq3}
\ee
where the specific enthalpy is defined by
\be
h = 1+\epsilon + \frac{P}{\rho_0}.\label{eq4}
\ee
The active mass density is then given by
\be
\rho_{\rm{active}} = T - 2T^{0}_{0} = T^{i}_{i} - T^{0}_{0} = \rho_0 h \frac{1+v^2}{1-v^2} +2P. \label{eq5}
\ee
In \refeqns{eq3}, \eqref{eq4} and \eqref{eq5}, $\rho_0$ is the rest mass density which is
proportional to the number density of baryons of the fluid, $P$ is the pressure, $u^\mu$ is the four
velocity of a fluid element with respect to an Eulerian observer, $\epsilon$ is the specific internal
energy, $T$ is the trace of the energy-momentum
tensor ($T=g_{\mu\nu}T^{\mu\nu}$), and $v$ is the three dimensional fluid velocity.
Unlike the Newtonian case, $\rho_{\rm{active}}$ includes all sources of energy.

The equations governing the motion of the fluid matter can be derived from
the conservations laws for the energy-momentum tensor and the fluid's matter current,
i.e., $\nabla_\mu T^{\mu\nu} = 0$ and $\nabla_\mu J^\mu = 0$.
In the ADM decomposition of spacetime \citep{ADM}, the metric ($g_{\mu\nu}$) can be expressed in the
following form by considering the foliation of spacetime using 3-dimensional hypersurfaces defined
by $t={\rm const.}$:
\be
ds^2=-\alpha^2 dt^2 + \gamma_{ij}\left(dx^i+\beta^i dt\right)\left(dx^j+\beta^j dt\right),\label{eq6}
\ee
Here
$\gamma_{ij}$ is the spatial metric, defined on each hypersurface,  while
$\alpha$ and $\beta^i$ are known as the lapse, and shift vector, respectively, and encode the
4-fold coordinate freedom of general relativity.

Flux-conservative formulations of hydrodynamics have been applied very successfully in
computational fluid dynamics.
To cast the fluid equations in flux-conservative form we first
define so-called conservative variables ($q$) in terms of the original hydrodynamic
variables (so-called primitive variables, $w$),
\be
q=\left(
\begin{array}{c}
D \\ S_i \\ \tau
\end{array}
\right)
=\left(
\begin{array}{c}
\rho_0 W \\ \rho_0 h W^2 v_i \\ \rho_0 h W^2 -P -D
\end{array}
\right),\,
w=\left(
\begin{array}{c}
\rho_0 \\ v^i \\ P
\end{array}
\right),
\label{eq7}
\ee
where $W=1/\sqrt{1-\gamma_{ij}v^iv^j}$.
With these definitions, and with the metric~(\refeqn{eq6}), we can then write
the Euler equation as \citep{fon00}
\be
\frac{\partial\left(\sqrt{\gamma}q\right)}{\partial t}
+\frac{\partial\left(\sqrt{-g}f^i\right)}{\partial x^i}
=\sqrt{-g}\Sigma, \label{eq8}
\ee
where the fluxes $f^i$ and the sources $\Sigma$  are given by
\ba
f^i&=&\left[
\begin{array}{ccccc}
D\left(v^i-\frac{\beta^i}{\alpha}\right)\\S_j\left(v^i-\frac{\beta^i}{\alpha}\right)+P\delta^{i}_{j}\\\tau \left(v^i-\frac{\beta^i}{\alpha}\right)+Pv^i
\end{array}
\right],\nonumber\\
\Sigma&=&\left[
\begin{array}{ccccc}
0\\
T^{\mu\nu}\left(\partial_{\mu}g_{\mu j}-\Gamma^{\lambda}_{\mu\nu}g_{\lambda j}\right)\\
\alpha\left(T^{\mu 0}\partial_{\mu}\left(\ln\alpha\right)-\Gamma^{0}_{\mu\nu}T^{\mu\nu}\right)
\end{array}
\right].\label{eq9}
\ea
Here $\sqrt{\gamma}$ and $\sqrt{-g}$ are the determinants of $\gamma_{ij}$ and $g_{\mu\nu}$,
respectively, and are related by $\sqrt{-g}=\alpha\sqrt{\gamma}$.
It is well known that for a perfect fluid, the system of equations derived from the
conservation laws is not closed:  the number of dynamical equations is always
less than the number of unknowns.

As is also well known, the equation of state (hereafter EOS) for the fluid  provides
an additional equation, but in the general case it also introduces other unknowns. In order
to completely close the hydrodynamical equations, an energy balance equation is often used. However,
under certain circumstances, we can adopt rather simple EOSs that do not
introduce any further variables: adiabatic and isothermal EOSs provide
specific examples.

Realistic EOSs are usually determined by theoretical calculations and experimental measurements.
However, there are physical regimes where our understanding of the nature of the matter is
quite incomplete.  Specifically, in the case where the matter density is significantly above
nucleon density, there remain large uncertainties in the correct EOS.
Thus, for example, the EOS at the core of neutron stars is still not very well understood.
Here we ignore these difficulties, and for the purpose of testing our code, use two
types of very simple EOS.
The first is the ideal gas EOS which can be written in the following form:
\be\label{eq30}
P=\left(\Gamma-1\right)\rho_0\epsilon,
\ee
and corresponds to the isothermal EOS.
We use this EOS in the shock tube tests described in (section \ref{testshtb}).
The second EOS results from the isentropic assumption, whereby \refeqn{eq30} becomes
the polytropic EOS:
\be\label{eq31}
P=K\rho_{0}^{1+\frac{1}{N}}.
\ee
Here $K$ and $N$ are the polytropic constant and index respectively.
The polytropic EOS of state is the generalized form of the adiabatic one;
a fluid which is governed by it does not generate entropy, and shock formation is
thus generically prohibited.
We use this EOS in the pulsation mode test (section \ref{testradmode} and
Appendix \ref{radpuleq}).

Using the above formulation, we are now ready to describe in detail the pseudo-Newtonian hydrodynamical
equations used in our code.
We limit our study here to axisymmetric systems, and adopt cylindrical coordinates
($R,Z,\phi$) such that
\be
ds^2 = -(1+2\Phi)dt^2 + \frac{1}{1+2\Phi}\left(dR^2+dZ^2+R^2 d\phi^2\right).\label{eq10}
\ee
The lapse function and shift vector are thus given  by $\alpha=\sqrt{1+2\Phi}$ and $\beta^i=0$.
In addition, we enforce the equatorial symmetry at $z=0$ since the phenomena involving $l=\rm{odd}$ modes are not dominant in most cases of neutron star dynamics, where $l$ is from the spherical harmonics.
In this coordinate system, the conservative and primitive variables are
\be
q=\left(
\begin{array}{c}
D \\ S_R \\ S_Z \\ S_{\phi} \\ \tau
\end{array}
\right)
=\left(
\begin{array}{c}
\rho_0 W \\ \rho_0 h W^2 v_R \\ \rho_0 h W^2 v_Z \\ \rho_0 h W^2 v_{\phi} \\ \rho_0 h W^2 -P -D
\end{array}
\right),\,
w=\left(
\begin{array}{c}
\rho_0 \\ v^R \\ v^Z \\ v^{\phi} \\ P
\end{array}
\right).
\label{eq11}
\ee
The final form of the hydrodynamical equations then becomes
\be
\frac{\partial\left(\sqrt{\gamma}q\right)}{\partial t}
+\frac{\partial\left(\sqrt{-g}f^R\right)}{\partial R}
+\frac{\partial\left(\sqrt{-g}f^Z\right)}{\partial Z}
=\sqrt{-g}\Sigma
\label{eq13}
\ee
where
\ba
f^R&=&\left[
\begin{array}{ccccc}
Dv^R\\S_Rv^R+P\\S_Zv^R\\S_{\phi}v^R\\\tau v^R+Pv^R
\end{array}
\right],\nonumber\\
f^Z&=&\left[
\begin{array}{ccccc}
Dv^Z\\S_Rv^Z\\S_Zv^Z+P\\S_{\phi}v^Z\\\tau v^Z+Pv^Z
\end{array}
\right],\nonumber\\
\Sigma&=&\left[
\begin{array}{ccccc}
0\\-\frac{\rho_{\rm{active}}}{1+2\Phi}\frac{\partial\Phi}{\partial R}+\frac{S_{\phi}v^{\phi}}{R}+\frac{P}{R} \label{eq14-1}\\
-\frac{\rho_{\rm{active}}}{1+2\Phi}\frac{\partial\Phi}{\partial Z}\label{eq14-2}\\
0\\-\left(S_R\frac{\partial\Phi}{\partial R}+S_Z\frac{\partial\Phi}{\partial Z}\label{eq14-3} \right)
\end{array}
\right]. \label{eq14}
\ea
Using \refeqn{eq10} we have $\sqrt{\gamma}=R\left(1+2\Phi\right)^{-3/2}$, and
$\sqrt{g}=R\left(1+2\Phi\right)^{-1}$.
In obtaining the expressions in \refeqn{eq14} we have used the assumption of slow changes of the
potential relative to the gradients ($\pdrv{\Phi}{t} \ll \pdrv{\Phi}{R}$ or $\pdrv{\Phi}{Z}$).
Recently, \cite{nag11} used a similar method in the context of jet propagation in a uniform medium,
but adopted a slightly different linear momentum equation than ours. (See \refeqn{eq14} and
compare with Eqs. (2) and (3) in \citealp{nag11}).

Finally, the gravitational Poisson equation in our coordinate system is
\be
\frac{1}{R}\frac{\partial}{\partial R}\left(R\frac{\partial \Phi}{\partial R}\right) + \pdrvs{\Phi}
{Z}=4\pi\rho_{\rm{active}}.
\label{eq15}
\ee

Note that  the second component of $\Sigma$ contains terms which, individually, become singular on the axis of
symmetry ($R=0$).  In addition, there are other terms in the equations of motion that need to be
treated carefully as $R\to0$.  This is done by demanding regularity at the axis, and by considering the
parity of each function, with respect to $R$, in that limit.
In particular, $\rho_0$, $v^Z$, $v^{\phi}$, $P$, $D$, $S_Z$, $S_{\phi}$, and $\tau$ are all even functions
of $R$ as $R\to0$, while $v^R$ and $S_R$ are odd.
Taking this into account,
\refeqn{eq13} and $\Sigma$ in \refeqn{eq14} become
\be
\frac{\partial\left(\sqrt{\gamma'}q\right)}{\partial t}
+2\frac{\partial\left(\sqrt{-g'}f^R\right)}{\partial R}
+\frac{\partial\left(\sqrt{-g'}f^Z\right)}{\partial Z}
=\sqrt{-g'}\Sigma
\label{eq16}
\ee
and
\be
\Sigma=\left[
\begin{array}{ccccc}
0\\0\\-\frac{\rho_{\rm{active}}}{1+2\Phi}\frac{\partial\Phi}{\partial Z}\\
0\\-S_Z\frac{\partial\Phi}{\partial Z}
\end{array}
\right],
\label{eq17}
\ee
where $\sqrt{\gamma'}=\left(1+2\Phi\right)^{-3/2}$ and $\sqrt{-g'}=\left(1+2\Phi\right)^{-1}$.
The coefficient of $\pdrv{\left(\sqrt{-g}f^R \right)}{R}$ in \refeqn{eq16} becomes $2$ instead of
$1$, while the other variables, such as $q$, $f^R$ and $f^Z$, are unchanged from \refeqn{eq14}.

Finally, using L'Hopital's theorem at $R=0$,
the singular term $R^{-1}\partial\Phi/\partial R$
in the Poisson equation~(\ref{eq15}) is replaced by
$\partial^2\Phi/\partial R^2$.


\section{Numerical Methods}\label{numer}
In this section, we describe our numerical methods for solving the coupled
hydrodynamical and Poisson
equations. We mainly use the finite volume methods for the hydrodynamical equations and the
a finite difference approach for the Poisson equation.
In the finite volume method, each grid cell represents volume averaged hydrodynamic quantities i.e.,
$\bar{q}=\frac{1}{\Delta V}\int q dV$.
After applying the finite volume method our hydro equations can be reduced to Riemann problems which
consider the time evolution of initial conditions given by two distinct states
that adjoin at some interface (so that there are, in general, discontinuities across one or
more physical quantities at the interface).
A very important property of the finite volume method is that it maintains the local conservation
properties of the flow in the computational grid.

In the dynamics of compressible fluids, we inevitably encounter discontinuous behaviors such as shocks,
rarefactions or contact discontinuities.
To treat such discontinuities without introducing numerical instabilities or spurious oscillations,
we use High Resolution Shock Capturing
(HRSC) techniques that generically reduce the order of accuracy of the numerical scheme near discontinuities
or when one or more of the  fluid variables are at a local maximum.
A key ingredient to the success of the HRSC methods is the calculation of fluxes through
cell boundaries.
To compute these fluxes we need approximate values for the primitive variables at the cell boundaries.
We have implemented second order slope limiters such as minmod \citep{van79}, monotonized central
difference (MC hereafter, \citealp{van77}) and superbee \citep{roe85}, as well as
a third order slope limiter  proposed by \cite{shi03} and which is based on the minmod
function (3minmod
hereafter). Other reconstruction methods such as the third order Piecewise Parabolic Method (PPM hereafter,
\citealp{col84}), Essentially Non-Oscillatory method (ENO,
\citealp{har87}) and Weighted ENO (WENO, \citealp{liu94}; \citealp{jia96}), which has an arbitrary order of
accuracy, were also implemented.

In the implementation of HRSC schemes it
is not efficient to exactly solve the Riemann problems which arise since an excessively large
amount of computational resources per cell are then needed
to calculate the fluxes.
Thus, an approximate calculation of fluxes is performed.
We implemented the following three schemes: Roe \citep{roe81}, Marquina
\citep{don96,don98}, and HLLE \citep{har83,ein88,ein91} approximations.
The Roe approximation is based on the Rankine-Hugoniot jump condition  and Marquina's approach
generalizes Roe's scheme.
The HLLE algorithm comes from a very simple two wave approximation and produces the most dissipative and stable
results.
We have mainly used HLLE for reducing computational cost but found that our results were not significantly influenced by the type of
flux approximation used.


In order to solve the Poisson equation (which is elliptic) for the gravitational potential we use
the multigrid method, which can quickly reduce low frequency error components
in the solution by adopting hierarchical grid
 levels \citep{bra77}. One of the difficulties we often encounter with the Poisson equation is in the
proper implementation of the boundary conditions. For example, one of the natural boundary conditions
is $\Phi =0$ at $\infty$, but in the coordinates adopted in the previous section the computational domain
cannot reach
spatial infinity.
We thus now refer to our previous coordinates as $(r,z)$ and introduce new coordinates $(R,Z)$ which
compactify the spatial domain, mapping the infinities
in each spatial direction to finite coordinate values.
Specifically,
we choose the same type of compactification for both $r$ and $z$ coordinates, namely a tangent function,
but allow a certain portion of the domain to remain '''uncompactified'':
\ba
r=
\left\{
\begin{array}{ll}
R & \mbox{if $R \leq r_0$} \\
r_0+r_1\tan\left(\frac{R-r_0}{r_1}\right) & \mbox{if $R > r_0$}
\end{array}
\right. , \label{eq21}\\
z=
\left\{
\begin{array}{ll}
Z & \mbox{if $Z \leq z_0$} \\
z_0+z_1\tan\left(\frac{Z-z_0}{z_1}\right) & \mbox{if $Z > z_0$}
\end{array}
\right. .\label{eq22}
\ea
Here, the four parameters $z_0$, $z_1$, $r_0$ and $r_1$ control the compactification,
and we chose  this specific form for the coordinate transformation since it
guarantees that the compactified coordinates smoothly transition to  the original
ones near the origin.
We note that we solve the hydrodynamical and gravitational equations on separate spatial
domains:
$[0:r_0,0:z_0]$ for the hydrodynamic calculations and $[0:r_0+\frac{2}{\pi}r_1,0:z_0+\frac{2}{\pi}z_1]$
for the computation of the gravitational potential. There ranges correspond to $[0:r_0,0:z_0]$ and
$[0:\infty,0:\infty]$, respectively, in the original cylindrical coordinates $(r,z)$.
In the compactified coordinates, the Poisson equation is written as
\ba
&&\frac{1}{r f(R)}\frac{\partial}{\partial R}\left(\frac{r}{f(R)}\frac{\partial \Phi(R,Z)}{\partial R}\right)  \nonumber \\
&&+\frac{1}{g(Z)}\frac{\partial}{\partial Z}\left(\frac{1}{g(Z)}\frac{\partial \Phi(R,Z)}{\partial Z}\right)
=4\pi\rho_{\rm{active}},  \label{eq25}
\ea
where $f(R)$ and $g(Z)$ are given by
\ba
f(R)=\frac{dr}{d R}=
\left\{
\begin{array}{ll}
1 & \mbox{if $R \leq r_0$} \\
\sec^2\left(\frac{R-r_0}{r_1}\right) & \mbox{if $R > r_0$},
\end{array}
\right.\label{eq26}\\
g(Z)=\frac{dz}{d Z}=
\left\{
\begin{array}{ll}
1 & \mbox{if $Z \leq z_0$} \\
\sec^2\left(\frac{Z-z_0}{z_1}\right) & \mbox{if $Z > z_0$}
\end{array}
\right. .\label{eq27}
\ea
As just noted, the domain for the hydrodynamical calculation is finite,
i.e.~we do not solve the hydrodynamical equations
on the full compactified domain,
and we thus must be careful to choose values of $r_0$ and $z_0$ large enough so that there
is no outflux of matter through the $r=r_0$ and/or $z=z_0$ boundaries.
In our code we set $r_0=z_0=\eta r_e$ where $\eta$ is a free parameter and $r_e$ is the equatorial radius of
the rotating star as obtained from the procedure we use to calculate the initial stellar model.
For the pulsation mode test described in section \ref{testradmode}, a typical choice is $\eta=2$.
This means that the hydrodynamical computational domain extends twice the distance of the stellar
radius in
both the $R$ and $Z$ directions: this choice is found to be sufficient for our study.
The values of $r_1$ and $z_1$ are automatically determined by requiring the multigrid domain to be
2 times larger than the size of hydrodynamic domain in compactified coordinates,
i.e., $r_1=\frac{\pi}{2}r_0$ and $z_1=\frac{\pi}{2}z_0$.

In our multigrid algorithm, we use line relaxation for our basic smoother, whereby all grid point
values given by $R={\rm const.}$ or $Z={\rm const.}$ are updated simultaneously (constant-$R$
and constant-$Z$ sweeps are alternated).  We cannot use
point-wise relaxation since, as is well known, such a technique is not a good smoother when
there is significant anisotropy in the coefficients of the second derivative terms in the
elliptic operator being treated. This is the case in our compactified coordinate system, particularly
near the domain boundaries.
We have used second- and fourth-order finite-difference approximations to the Poisson equation,
and these lead to tri- and pentadiagonal linear systems, respectively, that must be solved
to implement the line relaxations.  We use the routines DGTSV (tridiagonal) and
DGBSV (banded/pentadiagonal) routines from LAPACK to perform these solutions.

In order to integrate the discretized hydrodynamical equations, we use the method of lines (MOL),
transforming our partial
differential equations in time and space to ordinary differential equations (ODEs) with respect to the time.
To solve these ODEs, we then employ second and third order Runge-Kutta methods, which are known to have the
Total Variation Diminishing (TVD) property.

\section{Shock tube Tests}\label{testshtb}
In order to verify the accuracy and the convergence of our numerical code,
we first carry out rigorous tests using initial configurations having analytic
solutions. In this section, we present the results of such tests for the case
where there is no self-gravity (i.e.~pure hydrodynamics).
Another test of the entire code---including our treatment of the gravitational field---is
described in the next section.

Shock tube tests are Riemann problems where the initial configuration of the fluid is
given by two states having, in general,
different densities, pressures and velocities,
on the left and right halves of the tube.  Three possible distinct features emerge
from the subsequent evolution: a shock, a rarefaction fan, and a contact discontinuity.
We carried out 1D and 2D numerical simulations with 3 different parameter sets previously
used by \cite{zha06}. These parameters are listed in Table \ref{tb1}, where
superscripts $R$ and $L$ represent the fluid states in the right and left halves,
respectively, of the tube.

\begin{table}
\begin{center}
\caption{Initial values of physical quantities for Shock tube tests (Riemann problem).}
\begin{tabular}{cccccccc}
\hline\hline
Problem & $\Gamma$ & $\rho_0^{L}$ & $\rho_0^{R}$ & $v^{L}$ & $v^{R}$ & $P^{L}$ & $P^{R}$ \\
\hline
1 & 5/3 & $10.0$ & $1.0$ & $0.0$ & $0.0$ & $13.33$ & $10^{-8}$ \\
2 & 5/3 & $1.0$ & $1.0$ & $0.0$ & $0.0$ & $1000.0$ & $10^{-2}$ \\
3 & 4/3 & $1.0$ & $1.0$ & $0.9$ & $0.0$ & $1.0$ & $10.0$ \\
\hline\hline
\end{tabular}
\end{center}\label{tb1}
\end{table}

\subsection{1D test in  Cartesian Coordinates}\label{1dtestshtb}
We first we present the results of our 1D tests.
In these tests, we carefully examine how the distinct features predicted by the
analytic solutions are reproduced by different methods, and measure the accuracy and
convergence rate of the various solutions obtained.

In problem 1, the initial discontinuity gives 3 different types of solutions
(shock, rarefaction, contact discontinuity).
Figure 1 shows the results at $t=0.4$ obtained using four different methods of reconstruction:
 minmod (top left), MC (top right), 3minmod (bottom left) and PPM (bottom right).
We observe that the minmod method is quite dissipative, yielding rather smooth solutions
that cannot accurately describe the shockwave.
We also find that at low resolution the height of the shock is not well reproduced
if we use the minmod method.  MC and 3minmod give almost similar results, while
PPM shows the best behaviour near the shock.

\begin{figure*}
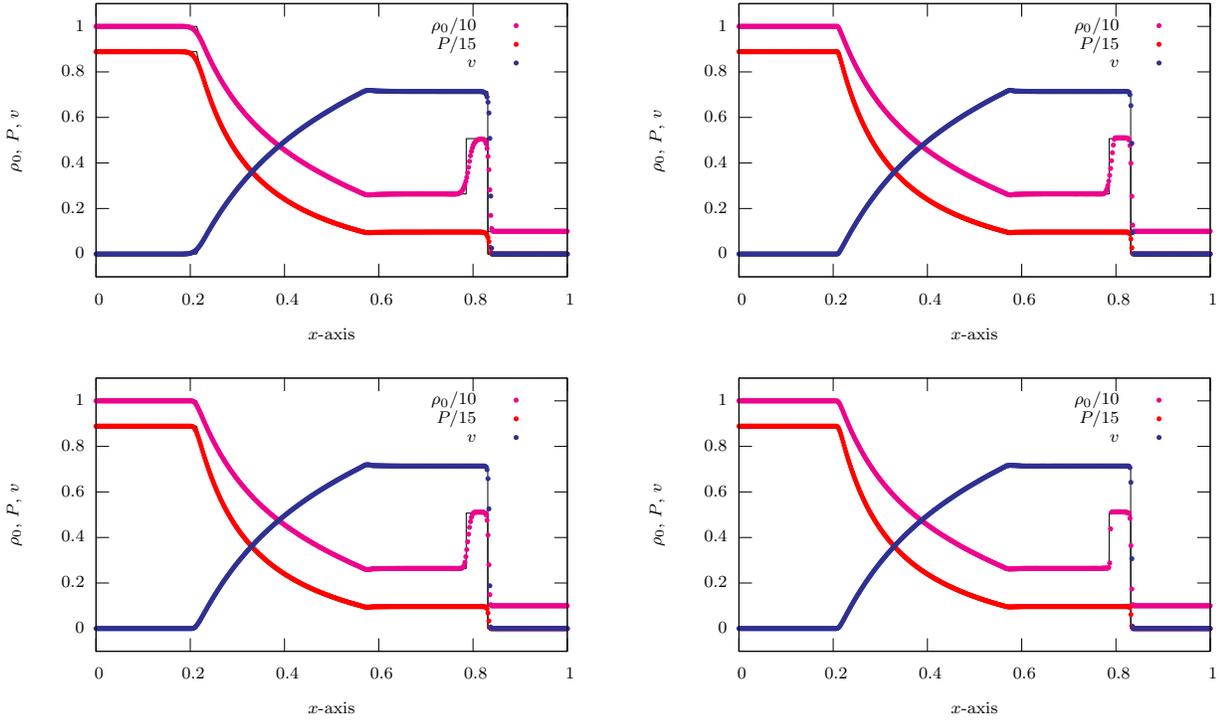
\label{fig01}
\begin{center}
\begin{tabular}{cc}
\scalebox{0.8}{\input{figure/shocktube1_minmod.tex}}&
\scalebox{0.8}{\input{figure/shocktube1_MC.tex}}\\
\scalebox{0.8}{\input{figure/shocktube1_shibata.tex}}&
\scalebox{0.8}{\input{figure/shocktube1_PPM.tex}}\\
\end{tabular}
\caption{One dimensional shock tube test of problem 1 at $t=0.4$ with different reconstruction methods:
minmod(top-left), MC (top right), 3minmod (bottom left), and PPM (bottom right). The initial
discontinuity is at $x=0.5$. We use 512 uniform grid points. The numerical results are shown in 3
different colours: rest mass density (pink), pressure (red) and velocity (blue). The solid lines show
the analytic solutions.}
\end{center}
\end{figure*}

\begin{figure*}
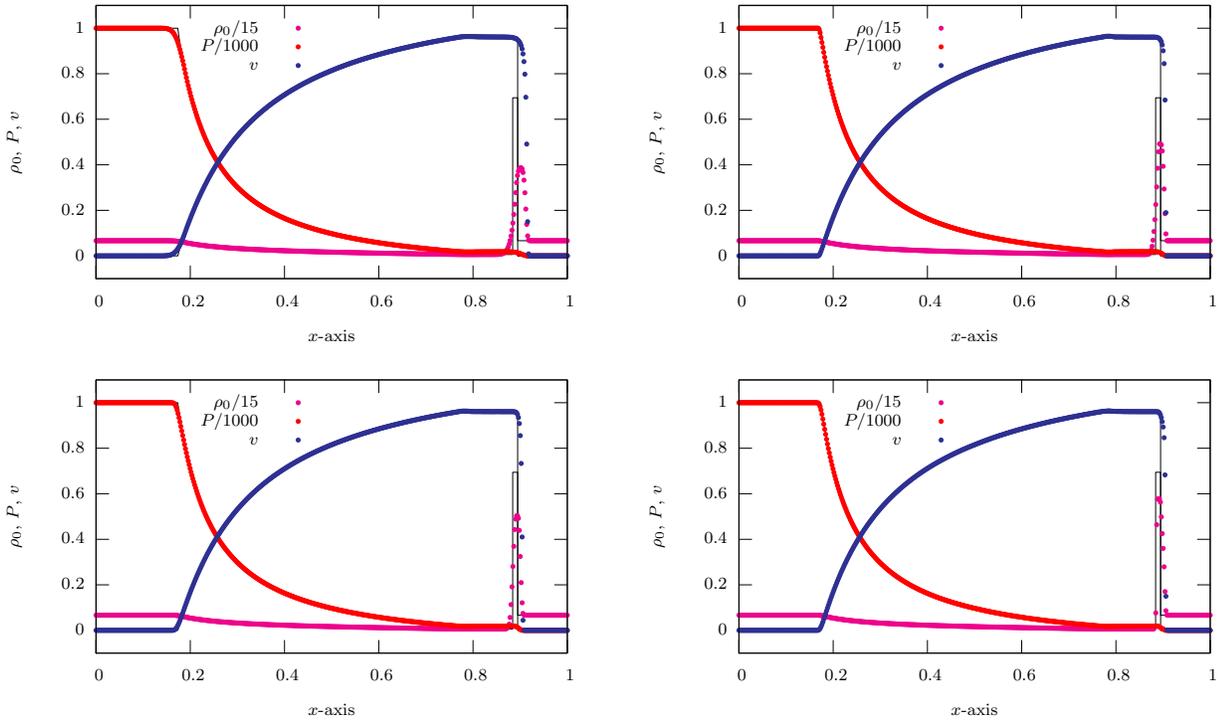

\begin{center}
\begin{tabular}{cc}
\scalebox{0.8}{\input{figure/shocktube2_minmod.tex}}&
\scalebox{0.8}{\input{figure/shocktube2_MC.tex}}\\
\scalebox{0.8}{\input{figure/shocktube2_shibata.tex}}&
\scalebox{0.8}{\input{figure/shocktube2_PPM.tex}}\\
\end{tabular}
\caption{Same as Fig. 1 for problem 2.}
\end{center}\label{fig02}
\end{figure*}

The second test problem (Problem 2) is the so-called blast wave test which produces a very sharp and thin
shell in density between the shock and contact discontinuity.
Generally, numerical codes are not able to perfectly resolve this very thin shell because it can span
only a few grid cells, even in very high resolution calculations.
Nonetheless, this test provides insight as to how well a code can handle such a feature.
As can see in Figure \ref{fig02}, PPM again gives the best results.
although it still shows large errors at the shock.

\begin{figure*}
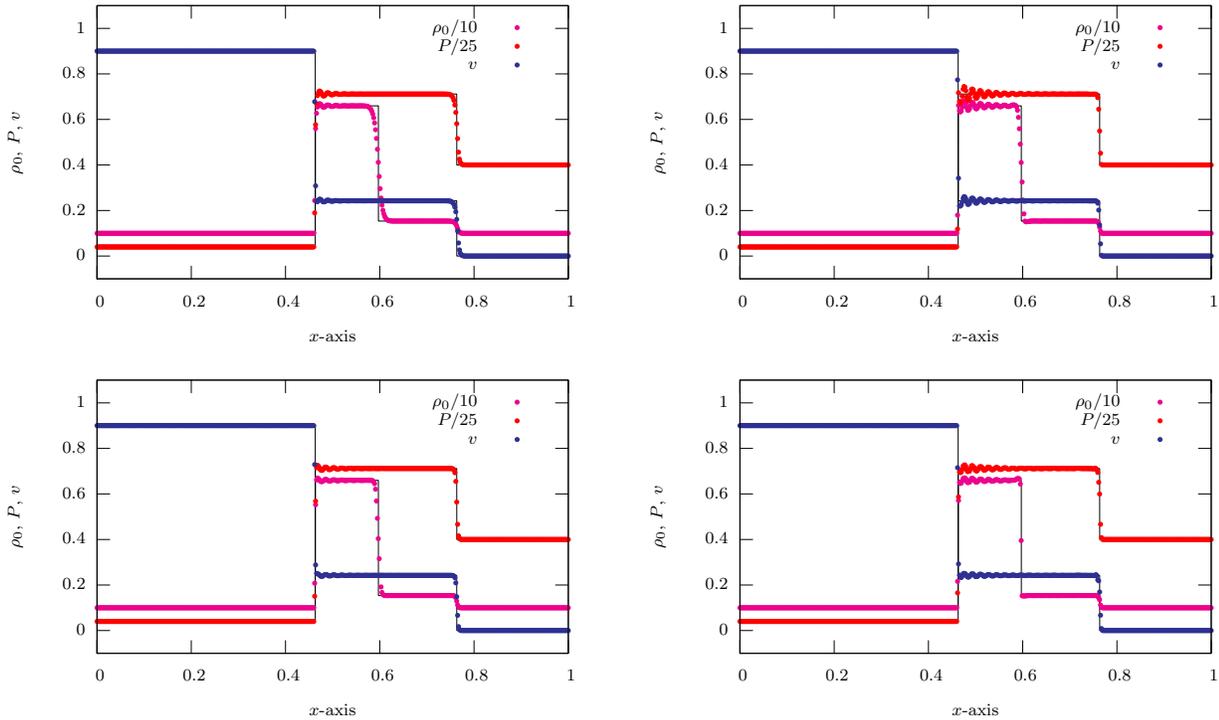

\begin{center}
\begin{tabular}{cc}
\scalebox{0.8}{\input{figure/shocktube3_minmod.tex}}&
\scalebox{0.8}{\input{figure/shocktube3_MC.tex}}\\
\scalebox{0.8}{\input{figure/shocktube3_shibata.tex}}&
\scalebox{0.8}{\input{figure/shocktube3_PPM.tex}}\\
\end{tabular}
\caption{Same as Figs. 1 and 2 for problem 3.}
\end{center}\label{fig03}
\end{figure*}

The third problem generates a strong reverse shock but numerical solution has oscillatory features near the shock front.
Generally speaking, the oscillation can be easily damped out if the numerical scheme is
significantly dissipative.
Numerical dissipation also tends to weaken the sharpness of the discontinuity.
In Figure \ref{fig03}, one can see that the minmod methods, which, as already noted, is the most dissipative
of the techniques we use, gives relatively small amplitude
oscillations, except near the discontinuity. The more non-dissipative methods describes
the shock features well, but produce rather large amplitude oscillatory behavior.

To quantify the deviation of our numerical results from the analytic solutions, we use the $L_1$ norm of
the errors,
defined by $L_1=\sum_{i=1}^{N}\Delta x_i |q_i-q(x_i)|$, where $q(x_i)$ is the value of the
analytic solution at point $x_i$.
We summarize the results in Table \ref{tb2}.
The convergence rate ($\log_2 \left[L_1^{2h}/L_1^{h}\right]$) in the table should be close to 1, which corresponds to the $1^\textrm{st}$ order nature of the HRSC scheme near the shock where the most of the $L_1$-norm error occurs.
However, it can deviate from that value due to the oscillatory features near the shock.

\begin{table*}
\begin{center}
\caption{The $L_1$ norm of the error and its convergence rate for each of the test problems using different resolutions and different reconstruction schemes.}
\begin{tabular}{ccccccccc}
\hline\hline
\multicolumn{3}{c}{} & \multicolumn{6}{c}{N} \\
\multicolumn{3}{c}{} & 64 & 128 & 256 & 512 & 1024 & 2048 \\
\hline
\multirow{8}{*}{Problem 1}&\multirow{2}{*}{minmod} & $L_1$ norm ($\times 10^{-2}$) & $25.7$ & $16.5$ & $9.48$ & $5.02$ & $2.66$ & $1.49$ \\
                          &                        & convergence rate              &    -   & $0.64$ & $0.80$ & $0.92$ & $0.91$ & $0.84$ \\ \cline{2-9}
                          &\multirow{2}{*}{MC}     & - & $15.3$ & $9.43$ & $5.25$ & $2.79$ & $1.46$ & $0.830$ \\
                          &                        & - &    -   & $0.70$ & $0.85$ & $0.91$ & $0.93$ & $0.82$ \\ \cline{2-9}
                          &\multirow{2}{*}{3minmod}& - & $17.8$ & $11.0$ & $5.82$ & $2.99$ & $1.51$ & $0.816$ \\
                          &                        & - &    -   & $0.69$ & $0.92$ & $0.96$ & $0.98$ & $0.89$ \\ \cline{2-9}
                          &\multirow{2}{*}{PPM}    & - & $12.3$ & $6.55$ & $3.43$ & $1.74$ & $0.877$ & $0.431$ \\
                          &                        & - &    -   & $0.91$ & $0.93$ & $0.98$ & $0.99$ & $1.0$ \\ \hline
\multirow{8}{*}{Problem 2}&\multirow{2}{*}{-} & - & $30.1$ & $21.0$ & $20.1$ & $15.8$ & $10.9$ & $6.93$ \\
                          &                   & - &    -   & $0.52$ & $0.061$ & $0.34$ & $0.54$ & $0.65$ \\ \cline{2-9}
                          &\multirow{2}{*}{-} & - & $27.8$ & $18.2$ & $14.9$ & $10.4$ & $6.28$ & $3.77$ \\
                          &                   & - &    -   & $0.61$ & $0.29$ & $0.52$ & $0.73$ & $0.74$ \\ \cline{2-9}
                          &\multirow{2}{*}{-} & - & $28.3$ & $17.9$ & $13.7$ & $8.84$ & $5.05$ & $2.72$ \\
                          &                   & - &    -   & $0.66$ & $0.39$ & $0.63$ & $0.81$ & $0.89$ \\ \cline{2-9}
                          &\multirow{2}{*}{-} & - & $29.5$ & $17.9$ & $12.7$ & $7.79$ & $3.73$ & $2.13$ \\
                          &                   & - &    -   & $0.73$ & $0.49$ & $0.71$ & $1.1$ & $0.81$ \\ \hline
\multirow{8}{*}{Problem 3}&\multirow{2}{*}{-} & - & $15.5$ & $10.0$ & $6.19$ & $3.65$ & $2.37$ & $1.58$ \\
                          &                   & - &    -   & $0.63$ & $0.69$ & $0.76$ & $0.63$ & $0.59$ \\ \cline{2-9}
                          &\multirow{2}{*}{-} & - & $14.9$ & $7.73$ & $5.40$ & $2.72$ & $1.64$ & $1.04$ \\
                          &                   & - &    -   & $0.95$ & $0.52$ & $0.99$ & $0.73$ & $0.66$ \\ \cline{2-9}
                          &\multirow{2}{*}{-} & - & $13.0$ & $6.97$ & $4.40$ & $2.25$ & $1.35$ & $0.867$ \\
                          &                   & - &    -   & $0.90$ & $0.66$ & $0.97$ & $0.74$ & $0.63$ \\ \cline{2-9}
                          &\multirow{2}{*}{-} & - & $7.26$ & $3.93$ & $2.41$ & $1.08$ & $0.547$ & $0.393$ \\
                          &                   & - &    -   & $0.89$ & $0.70$ & $1.16$ & $0.98$ & $0.47$ \\
\hline\hline
\end{tabular}
\end{center}\label{tb2}
\end{table*}

\begin{figure}
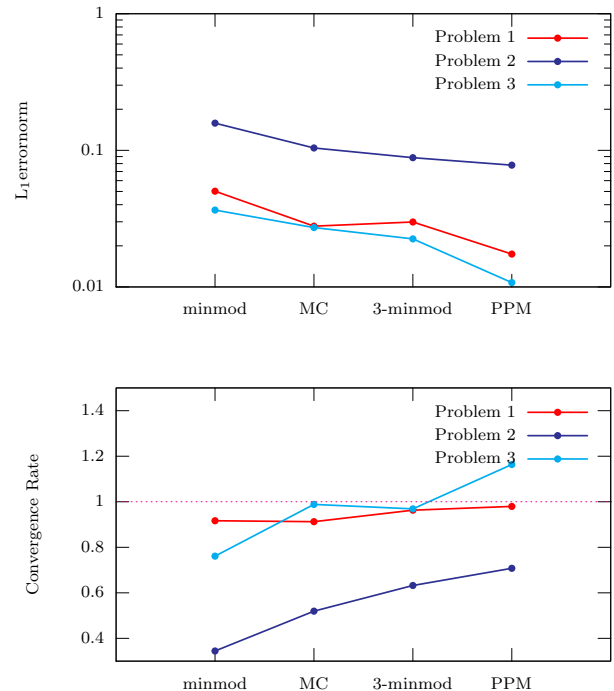

\begin{center}
\begin{tabular}{c}
\scalebox{0.8}{\input{figure/l1norm.tex}}\\
\scalebox{0.8}{\input{figure/conv_rate.tex}}
\end{tabular}
\caption{The $L_1$ norm (top) and convergence rate (bottom) when the number of grid points is 512 with different reconstruction methods. Three different shock tube problems are shown with different colors (problem 1: blue, problem 2: red and problem 3: sky blue).}
\label{fig04}
\end{center}
\end{figure}


Figure \ref{fig04} shows the $L_1$ norms and convergence rates for each problem when the grid
resolution is $\rm{N}=512$.
Although no single method stands out in our 1D shock tube tests, we
conclude from
from the values of the $L_1$ error norms and convergence rates
that PPM gives the most promising results.

\subsection{2D test in Cylindrical Coordinates}\label{2dtestshtb}
Since the cylindrical coordinate system we have adopted is curvilinear, 1-dimensional shock tube tests are
not sufficient for assessing our code's accuracy and convergence.
In  Cartesian coordinates, fluxes between cells which have the
same state cancel out. For example, if we carry out the shock tube test in the $x$-direction,
then the fluxes in the $y$ and $z$ directions are identical in every grid cell, meaning that the
net flux is 0. Therefore, 1D shock tube tests performed with codes that use
2- or 3D  Cartesian coordinates produce
exactly the same results as a 1D code.
However, in cylindrical coordinates, fluxes do not cancel in this way, but rather are balanced by
source terms.
This difference may give additional non-physical effects, especially near discontinuities.

Therefore, we carried out the first of the shock tube tests listed in Table~\ref{tb1} in cylindrical coordinates,
where we placed
the discontinuity on the Z=0 plane.
Figure 5 shows the solution resulting solution on the $Z$-axis.
If we use the minmod method, the 2D results are similar to the 1 dimensional ones.
In addition, although PPM produces better results minmod, it cannot produce
the sharp features of the shock seen in the 1D test: this is due to the dissipation caused by the
imbalance between the net flux and the source term.
We can also see that 3minmod and PPM yield quite similar results.
We checked the differences in solutions at different $z={\rm const.}$ planes and found that
they are negligibly small ($\sim 10^{-13}$) compared to the truncation errors.
Overall, however, although the 2D results show more dissipation than
the 1D ones, the relative differences in the solutions are
not significant (for our purposes). In particular, both agree acceptably with the analytic forms.

\begin{figure}
\begin{center}
\scalebox{0.8}{\input{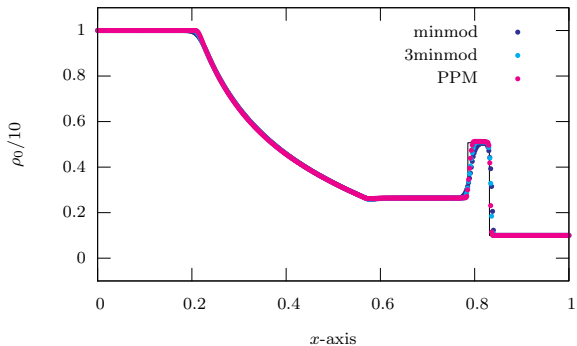}}
\caption{The solution on the axis at t=0.4 in problem 1 in section \ref{1dtestshtb}. The 3 different reconstruction methods (minmod: blue, 3-minmod: sky blue and PPM: magenta) are shown.}
\end{center}\label{fig05}
\end{figure}

\section{Stationary Star Test}\label{teststatstar}
The tests just reported did not involve the effects of the gravitational field.
In this section, we test our treatment of the Poisson equation
for the gravitational potential as well as the hydrodynamics.

With an ideal code, the evolution of a stationary star should also be stationary.
However, in practice, all codes that dynamically evolve stationary states show some level of
fluctuation due to finite grid resolution and
intrinsic errors in the numerical scheme used.
In this section, we show the time evolution of the physical quantities of non-rotating and the
rotating stars, and investigate the dependence of this time behaviour by changing the resolution of the
simulations.
Specifically, we use 3 different grid resolutions : $65\times65$, $129\times129$ and $257\times257$,
where $1/2$ of the grid points span the star at the equator.

Our initial models of  rotating stars are generated using  Hachisu's Self-Consistent Field
(HSCF: \citealp{hac86a,hac86b}) method---details of the procedure are described in \cite{kim09}.
In order to generate equilibrium models, we choose 1) the maximum rest mass
density, $\rho_0^{\rm{max}}$) 2) the rotation parameter, $A$, which describes the differential
rotation and 3) the axis ratio which determines how fast the star is rotating.
We must also specify the equation of state (EOS) in our construction of the initial model.
Here, we used the polytropic EOS~\refeqn{eq31} with $K=100$ and $N=1$.
We choose a  maximum density value of $\rho_0^{\rm{max}}=1.28\times10^{-3}$, which, with
this EOS,
produces a 1.4\solarmass star
in the non-rotating case.
For the rotating models, we only consider
rigid body rotation, which is
obtained when we chose a very large value of $A$.
The axis ratio is specified to be $0.75$ resulting in an orbital frequency of $611\rm{Hz}$.

Even with our use of the multigrid technique---which is generally an efficient method
for solving elliptic equations---we still find solution of the Poisson equation
for the gravitational potential to be computationally expensive.
We thus calculate $\Phi$ only every $40$ time steps to reduce the time spent in the Poisson solver,
and find that this produces results which are nearly equivalent to those obtained when the Poisson
equation is solved at each time step.
However, we use time-extrapolated values for the gravitational potential at the time steps
between solves of the Poisson equation
in order to avoid discontinuities in the primitive variables, when abrupt changes of the gravitational
potential occur. We find that these discontinuities give rise to very unnatural dissipative effects
in the simulation resulting, for example, in a rapid decay in the amplitude of radial oscillations,
even when radial perturbations are explicitly introduced.

\begin{figure*}
\begin{center}
\begin{tabular}{cc}
\scalebox{0.8}{\input{figure/static_cowling2.tex}}&
\scalebox{0.8}{\input{figure/static_rot_cowling2.tex}}\\
\scalebox{0.8}{\input{figure/static_dyn2.tex}}&
\scalebox{0.8}{\input{figure/static_rot_dyn2.tex}}
\end{tabular}
\caption{The time evolution of the maximum rest mass density changes($\left[\rho_0^{\rm{max}}
(t)-\rho_0^{\rm{max}}(t=0)\right]/\rho_0^{\rm{max}}(t=0)$) with different resolutions
($65\times65$: red, $129\times129$: dark blue and $257\times257$: sky blue).
Top figure shows results when we fix
the metric (Cowling approximation) while the bottom one shows the case where we consider the
fully coupled dynamics.
In the left panel, we show the figures for a spherical (non-rotating) star while the right panel
shows the corresponding figures for a rigidly rotating star with axis ratio$=0.75$, which
give a rotational frequency of
$611\rm{Hz}$}
\end{center}\label{fig06}
\end{figure*}

Figure 6 shows the time evolution of the relative changes of the maximum density
($\left[\rho_0^{\rm{max}}(t)-\rho_0^{\rm{max}}(0)\right]/ \rho_0^{\rm{max}}(0)$) for non-rotating (left
panel) and rigidly rotating (right panel) stars.
For the stationary stars, we use the Cowling approximation, which assumes the gravitational
potential is fixed. This gives efficient evolution of the stars, and can also be used as a testbed
for fully coupled evolutions.
The results computed using the Cowling approximation are shown in the top figures.
The maximum density slowly increases with time for the rotating star while it decreases for the
non-rotating star.
For grid resolutions greater than $65\times65$ the rate of change is almost independent of
resolution for the spherical star, but a slow decrease with resolution is
seen for the rotating star. We define the
following dimensionless rate of change:
\be
\mathcal{R}=\left|t_{\rm{dyn}}\frac{d \ln \rho_0^{\rm{max}}}{d t}\right|,\label{eq:lateofchange}
\ee
where we use $t_{\rm{dyn}}=1/\sqrt{\rho_0^{\rm{max}}}$ for simplicity.
We use this quantity---as computed from the highest resolution simulations---as a label in the figures.
The values of $\mathcal{R}$ are within $3\times10^{-7}$ for non-rotating star and are about 10 times larger
for the rotating star, again with a maximum resolution of $257\times257$.
The inverse of $\mathcal{R}$ can be interpreted as
the time (in units of the dynamical time) that the simulation could be carried out until the results
deviate from the true solution by $O(1)$.
Our results indicate that the error would become $\sim$ 1\% in 30,000 and 3,000
dynamical times for non-rotating and rotating stars, respectively.
We also carried out very long time simulations and found that $\mathcal{R}$ becomes smaller even
though it appears
to be almost constant in the figures. From these results, we conclude that we can use the
code to evolve stellar
configurations for several thousand or more dynamical times.

It is also very important to check the constancy of the conserved quantities with respect
to simulation time.
In our formulation we have 2 conserved quantities: the total rest mass, $M_0$,  and the total
angular momentum, $J$,
which are computing using
\ba
M_0=\int D dV^{(3)} = 2\pi\int \frac{\rho_0 W}{\left(1+2\Phi\right)^{3/2}}RdRdZ,\label{eq32}\\
J=\int S_{\phi} dV^{(3)} = 2\pi\int \frac{\rho_0 h W^2 v^{\phi}}{\left(1+2\Phi\right)^{5/2}}R^3dRdZ,
\label{eq33}
\ea
respectively, and
where $dV^{(3)}$ denotes the 3-dimensional volume element.
\begin{figure}
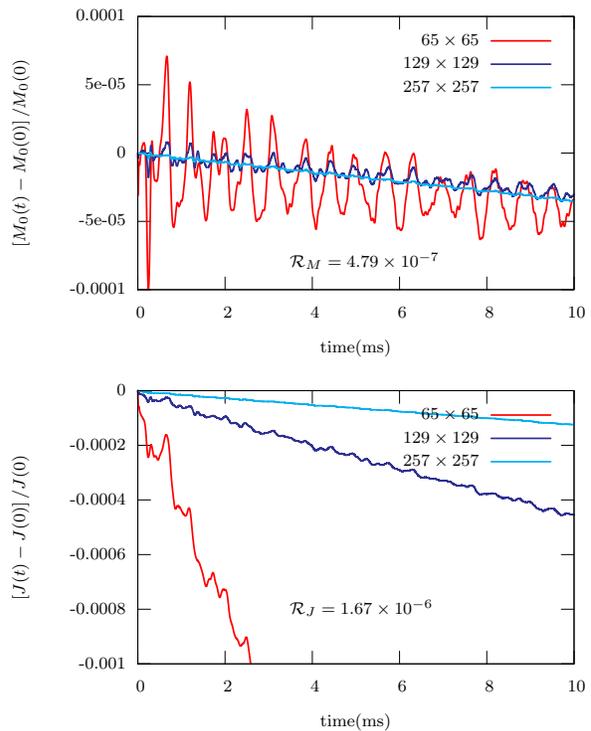

\begin{center}
\begin{tabular}{c}
\scalebox{0.8}{\input{figure/static_rot_dyn_M2.tex}}\\
\scalebox{0.8}{\input{figure/static_rot_dyn_J2.tex}}
\end{tabular}
\caption{The deviation of total rest mass (upper panel) and angular momentum (lower panel),
which should remain constant,
from their initial values with time, for the same models shown in Fig. 6.
Results computed with three different resolutions ($65\times65$:red,
$129\times129$:dark blue and $257\times257$:sky blue) are shown.}
\end{center}\label{fig7}
\end{figure}
Figure 7 shows the time evolution of these two conserved quantities:
total rest mass (upper panel) and total angular momentum (lower panel).
We show the results only from the rotating star since there is, of course, no angular momentum
for non-rotating stars. The deviation of the total rest mass from the initial value has two
features: short-term fluctuations and long term average behavior. The shot-term fluctuations
depend on the grid resolution, but the average slopes are almost independent of the
resolution. We label the graphs with $\mathcal{R}_M$ and $\mathcal{R}_J$ in a  manner analogous to
\refeqn{eq:lateofchange} and Fig.~\ref{fig06}, and use these quantities
to measure the long-term stability of the code. Their measured values are consistent with the ones for
the central density ($\mathcal{R}$). The behaviour of $\mathcal{R}_M$ is quite similar for the
3 different
grid resolutions, but $R_J$ shows considerable dependence on the grid resolution.
We have seen above (see Figure 6) that the
central density fluctuation is significantly dependent on grid resolution
only for the rotating models.
We conclude that the main reason for this
resolution-sensitive behavior
is the fact that angular momentum conservation is sensitive to
grid resolution. Therefore, simulations for rotating stars require high grid resolution,
otherwise angular momentum conservation will fail, and other stationary properties of the
start (such as central density) will also show substantial, and non-physical, time evolution.


\section{Radial Pulsation Frequency Test}\label{testradmode}
Even without any explicitly-added perturbations, it is natural for our numerical simulation
of stationary stars to give rise to normal
mode oscillations due to intrinsic numerical errors.
These errors occur for a variety of reasons, including 1) truncation error due to the
discretization scheme, 2) the artificial atmosphere
(floor) whereby the primitive variables (pressure, density) are restricted from falling
below minimum values to avoid code crashes (the sound velocity becomes
unbounded when vacuum is encountered in the numerical calculations), and
3) the numerical limitation in describing the
stellar surface. Furthermore, the artificial atmosphere is known to excite higher overtone modes.

The frequencies of various modes depend only on the structure of a given star, and can be calculated by
various methods. As explained above, our stationary models oscillate even when we do not
explicitly introduce external or internal perturbations.
We attempted to compare the frequencies of the modes
excited in our models with those obtained by normal mode analysis.
The fundamental mode (F-mode hereafter) frequency is very closely related to the dynamical time~($\sim
1/\sqrt{\rho}$) and the associated overtones have frequencies of similar order.

Although using calculations based on cylindrical coordinates is not an efficient way to
compute radial pulsations, our code
should still be able to approximately compute the correct pulsation frequencies.
The detailed perturbation formulations and numerical methods we use for
investigating the  radial pulsations are described in
Appendix \ref{radpuleq}. For initial conditions we use a non-rotating equilibrium star
with a baryon mass $1.4\solarmass$.
we performed the test with and without the Cowling approximation, and
In order to obtain the mode frequency from the simulations, we analyzed the fluctuation of
the maximum density with time.

Specifically, we carried out Fourier transformation on the maximum density using the FFTW package \citep{fri05}.
To obtain  better
resolution in the frequency domain, we use the zero-padding method which adds  additional zeros at the
end of the time series data,
effectively using interpolation between points following the basic Fourier transformations.
During the process of obtaining a frequency having a maximum sinusoidal amplitude,
leakage may also cause additional errors.
To reduce the effects of this leakage, we multiply the time series by a window function.
Here we used the Hamming window function defined by
\be\label{eq28}
w_j=0.54+0.46\cos\left(\frac{2\pi j}{N}\right),
\ee
where $j$ is the index of the grid points and $N$ is the total number of points,
prior to zero-padding \citep{har78}.

Although, as described above, some modes are excited simply due to numerical error, their
amplitudes are too small to be accurately extracted from the simulation.
We therefore introduce an explicit  perturbation which can more strongly excite the radial modes.
The perturbation that we used is
\be\label{eq29}
\delta\rho_0=B_s\sin\left(\pi \frac{r}{r_s}\right),
\ee
where $B_s$ is the perturbation amplitude which we set to $B_s=0.001$.

\begin{figure}
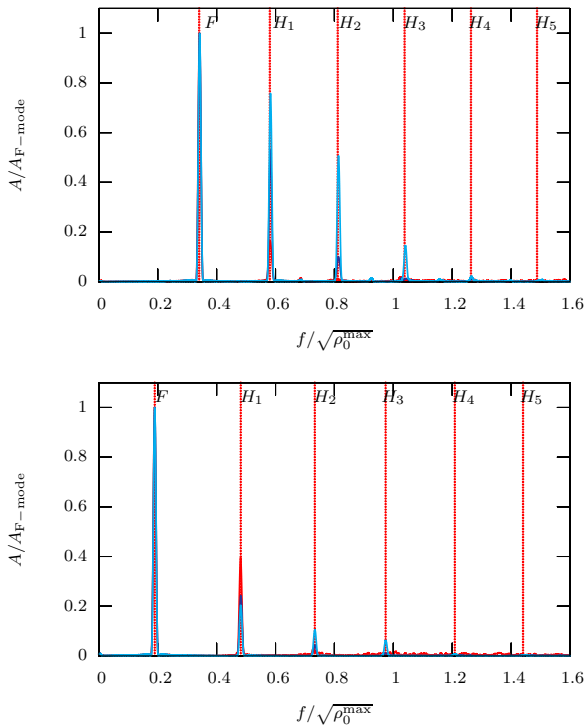

\begin{center}
\begin{tabular}{cc}
\scalebox{0.8}{\input{figure/compare_cowling.tex}}\\
\scalebox{0.8}{\input{figure/compare_dyn.tex}}
\end{tabular}
\caption{The mode amplitudes of maximum density as a function of frequency of the star with
a baryon mass $1.4\solarmass$. The vertical red dotted lines show the frequency of the
radial pulsation modes computed using the perturbation method. The top panel shows the result when we
use the Cowling approximation, where the gravitational potential is assumed to be fixed.
In the bottom figure, we obtain the gravitational potential every few time steps.
In the both panels the 3 curves show results obtained using 3 different
grid resolutions (sky blue: $257\times257$, dark
blue: $129\times129$, and red: $65\times65$)}
\end{center}
\label{fig08}
\end{figure}

Figure 8 shows the result after Fourier transformation of the time series data given by
the differences in maximum density relative to the
initial time ($\rho_0^{\rm{max}}(t)-\rho_0^{\rm{max}}(t=0)$),  and
using calculations at different resolutions.
For comparison purposes, the vertical red lines
show the results computed from linear analysis.
The mode labeled as $F$ is the fundamental mode, while $H_n$ denote the
$n$-th overtone radial modes. The results shown in the figure can be summarized as follows:
\begin{enumerate}
\item The most excited mode with the perturbation given by \refeqn{eq29} is the $F$-mode.
By changing the nature of the perturbation we could make one of the overtones the most highly
excited.
\item At low resolution, the code cannot identify high frequency modes. The reason for this is
the lack of spatial, not temporal, resolution. The eigenfunctions
describing  higher overtones
have large gradients near the surface which cannot be accurately represented in the low
resolution calculations.
\item The frequency increases when we use the Cowling approximation. This is a well-known phenomena
irrespective of whether Newtonian or general relativistic gravitation is used.
This issue is discussed in more detail in the appendix.
\end{enumerate}

Table \ref{tb3} shows the mode frequencies computed from linear analysis
as well as the numerical simulations.
Again, the stellar model is a non-rotating spherical star of mass 1.4 \solarmass. The
relative difference between the linear and full numerical results is listed in the
last row. Here the numerical simulations have been carried out using the highest resolution ($257\times257$),
and we list results computed
with and without the Cowling approximation.
\begin{table}
\begin{center}
\caption{Comparison of mode frequencies obtained by numerical simulation ($2\pi f$) and by
linear analysis ($\sigma$)
}
\begin{tabular}{c|ccccc}
\hline\hline
Mode & $F$ & $H_1$ & $H_2$ & $H_3$ \\
\hline
$f/\sqrt{\rho_0^{\rm{max}}}$           & $0.190$ & $0.482$ & $0.734$ & $0.974$ \\
\hline
$\sigma/2\pi/\sqrt{\rho_0^{\rm{max}}}$ & $0.190$ & $0.482$ & $0.733$ & $0.974$ \\
\hline
Error(\%)                                               & $0.000$  & $0.000$  & $0.136$ & $0.000$\\
$|2\pi f-\sigma|/\sigma$                                \\
\hline\hline
\end{tabular}
\end{center}\label{tb3}
\end{table}
The frequencies we obtained from the numerical simulation with $257\times257$ grid resolution
have relative differences from those computed from linear analysis of at most $0.1\%$.
We thus conclude that the radial mode frequencies computed from our code agree very well with the ones
calculated from linear theory. The largest difference of 0.1\% was found in the second overtone
(mode $H_2$), while for other modes we did not find any measurable difference.



\section{Summary and Discussion}\label{conclusion}
We have developed a new hydrodynamical code which adopts a pseudo-Newtonian treatment
of the gravitational field.
This code uses the so called ``Valencia formulation'' for the hydrodynamical equations.
From the computational perspective, the code is modular and includes many reconstruction schemes such as
slope limiting techniques (minmod, MC, 3rd order minmod, etc.), PPM and ENO (WENO). In 1D
shock tube tests, we assessed  code accuracy relative to analytic solutions and computed convergence
rates of the errors.
We found that the minmod method gives the most diffusive results,  smoothing out  complex features
near  discontinuities. As a result it cannot be used to accurately describe stellar surfaces, which are
characterized by stiff density changes. The MC method gives the most promising result in the shock tube test
and has second order accuracy. It can capture discontinuities very well in the pulsation mode test,
but also yields additional non-physical effects such as the excitation of the higher order overtones
near the stellar boundary. The 3minmod and PPM methods can provide higher order accuracy and we have
found that they can also describe the stellar surface well.

In the code we also implemented 3 different flux approximation schemes: Roe, Marquina,
and HLLE. Although the results in this paper were all computed using the HLLE approach---which
is the most dissipative of the three---we have also found that for the simulations we have
considered all produce very similar results.

In the multigrid module for computing the gravitational potential we have implemented both
second and fourth order finite-difference discretizations.
The actual value of the gravitational potential is slightly
different if we change the order of accuracy. However, the changes of maximum density in time
show very little sensitivity to the order of approximation, and we consider the difference
between the use of the second or fourth order method to be insignificant.

In the stationary star test which is described in section \ref{teststatstar}, we evolve equilibrium
solutions describing both non-rotating and rotating stars using our code.
Our code shows stable long-time behavior of the maximum density and conserved quantities.
Based on the rates of change in the maximum density, total mass and total angular momentum,
we estimate that our code can be used to study evolution in excess of 3,000 dynamical times with 1\% error.

In the radial mode test described in section \ref{testradmode}, modes are obtained from the Fourier
transformation of the maximum density fluctuations.
We also computed normal modes by linear analysis (see Appendix \ref{radpuleq}) and
found that the mode frequencies generated by our code agree with the results from linear
analysis almost perfectly (less than 0.14\%).


This code can be applied to the following astrophysical scenarios:
\begin{enumerate}
\item Phenomena associated with isolated rotating neutron stars, such as axisymmetric
pulsations.
    Since our approach can be applied to mildly compact stars, it is very useful to
    determine the amplitudes and frequencies of the radial and non-radial modes.

    \item Accretion disks around a neutron star or black hole.
It is not sufficient to treat a disk around a compact object using Newtonian gravity, since the
gravitational field is not weak there.
In addition, because the rotational velocity of the disk is a significant fraction of $c$,
we should also take into account special relativity in our treatment
of the hydrodynamics. Our code can be a very good tool for accretion disk studies.
\end{enumerate}

\vskip 3mm
This work was supported by the NRF grant 2006-341-C00018, by NSERC, and by the CIFAR
Cosmology and Gravity Program.

\begin{appendix}

\section{Perturbation equation}\label{radpuleq}
The eigenfrequencies and eigenfunctions of the radial pulsation of stars are well-known in Newtonian hydrodynamics as well as in the general relativistic case.
However, the corresponding  formulation has not been previously presented for
our pseudo-Newtonian approach.
Here, we describe the linearized equations that can be used to obtain eigenfrequencies and
eigenfunctions of the normal modes of spherical stars using this approximation, and
following the general relativistic framework described in \citealp{MTW} (MTW hereafter).
First, to describe stellar oscillations---such as those occurring on the  surface---it is  much
more practical to use a Lagrangian description rather than the Eulerian one adopted in
section \ref{formulation}.
The relation between the the Eulerian and Lagrangian perturbation is (see, e.g., Cox 1980),
\be
\Delta f(t,r) = \delta f + f_0' \zeta ,
\label{eqa3}
\ee
where $\zeta$ is a Lagrangian variation in space.
The law of baryon number conservation($\nabla_{\mu}(n u^{\mu})=0$) gives
\be
\Delta n = -n_{0}[r^{-2}\alpha_{0}^{3}(r^2\alpha_0^{-3}\zeta)'-3\alpha_0^{-1}\delta\alpha],
\label{eqa4}
\ee
where $\alpha=\sqrt{1+2\Phi}$, $n$ is the baryon number density and ${}'$ denotes  differentiation
with respect to $r$ (See MTW Eq.(26.7)).
The relation between $n$ in \refeqn{eqa4} and $\rho_{0}$ is $\rho_0 = m_b n$, where $m_b$ is
baryon mass and the subscript $0$ denotes the unperturbed state.

Another perturbation equation comes from the adiabatic equation of state which offers a much easier way
to find the pressure variation:
\be
\Gamma=\frac{n}{P}\frac{d P}{d n}.
\label{eqa13}
\ee
Since the Lagrangian variations commute with total differentiation (denoted by $d$),
\refeqn{eqa13} becomes
\be
\Gamma=\frac{n}{P}\frac{\Delta P}{\Delta n}.
\label{eqa5}
\ee
In addition, \refeqns{eqa3}, \eqref{eqa4} and \eqref{eqa5} give the following
pressure variation equation:
\be
\delta P = -\Gamma P_{0}[r^{-2}\alpha_{0}^{3}(r^2\alpha_0^{-3}\zeta)'-3\alpha_0^{-1}\delta\alpha]
-\zeta P_0'.
\label{eqa6}
\ee

The energy conservation equation ($u_{\mu}\nabla_{\nu}T^{\mu \nu}$) gives
\be
\Delta\rho=\frac{\rho_0 + P_0}{n_0} \Delta n .
\label{eqa7}
\ee
Note that $\rho_0$ is the energy density in the unperturbed state,
rather than the rest mass density used in the main text.
Combining this with \refeqn{eqa4}, we  obtain the equation for the energy density variation
\be
\delta \rho = -(\rho_0+P_{0})[r^{-2}\alpha_{0}^{3}(r^2\alpha_0^{-3}\zeta)'-3\alpha_0^{-1}\delta\alpha]
-\zeta \rho_0' .
\label{eqa8}
\ee

The main difference here relative to the general relativistic case arises in the computation
of the  perturbation of the
gravitational potential. The Poisson equation gives
\be
\frac{2}{r}(\alpha_0 \delta\alpha)'+(\alpha_0 \delta\alpha)''= 4\pi(\delta\rho+3\delta P) .
\label{eqa9}
\ee
Note that we should use only the Eulerian variation in this equation since Lagrangian variation does not
commute with partial differentiation.
Eq. (26.16) in MTW involves only first order differential equations---i.e. the second order
differentiations are rewritten in terms of first order ones.
On the other hand, in our case we cannot find any equations which can be used to eliminate
the second order differentiation.
That means that we need to find one more boundary condition to solve this equation.

Finally, the equation of motion of the fluid is obtained from the 4-acceleration
($a_{\mu}=u^{\nu}\nabla_{\nu}u_{\mu}$),
\begin{multline}
(\rho_0+P_0)\alpha_0^{-4}\ddot{\zeta}=-\delta P' -(\delta\rho+\delta
P)\alpha_0^{-1}\alpha_0'\\
-(\rho_0+P_0)(\alpha_0^{-1}\delta\alpha'-\alpha_0^{-2}\alpha_0'\delta\alpha).
\label{eqa10}
\end{multline}

Under the assumption of the adiabatic nature of the oscillation, normal modes are standing
waves, and thus space and time variables can be separated as follows:
\be
\zeta(r,t)=\xi(r)e^{i \sigma t}.
\ee
Then, we can rewrite the equations using $\xi$ and $\sigma$,
\ba
&&\delta P = -\Gamma P_{0}[r^{-2}\alpha_{0}^{3}
(r^2\alpha_0^{-3}\xi)'-3\alpha_0^{-1}\delta\alpha] -\xi
P_0' \label{eqa11-1}\\
&&\delta \rho = -(\rho_0+P_{0})[r^{-2}\alpha_{0}^{3}
(r^2\alpha_0^{-3}\xi)'-3\alpha_0^{-1}\delta\alpha]
-\xi \rho_0' \label{eqa11-2} \\
&&(\rho_0+P_0)\alpha_0^{-4}\sigma^2\zeta=\delta P' +(\delta\rho+\delta
P)\alpha_0^{-1}\alpha_0'\no
&&\qquad\qquad\qquad\qquad+(\rho_0+P_0)(\alpha_0^{-1}\delta\alpha'-\alpha_0^{-2}\alpha_0'\delta\alpha)
\label{eqa11}
\ea
To solve \refeqns{eqa11-1}--\eqref{eqa11}, we need to impose appropriate boundary
conditions.
The first condition is that $\xi/r$ should be regular at the origin, and the second one is
that the pressure variation at the surface must vanish, i.e.,
\ba
\frac{\xi}{r}={\rm finite~at~} r=0, \\
\Delta P(r=r_s)=0.
\ea
Unlike the general relativistic case, we cannot substitute $\delta\alpha$ and $\delta\alpha'$
in terms of other variations such as $\delta\rho$ and $\delta P$.
Therefore, we need an additional boundary condition for \refeqn{eqa9}.
We use the properties of the gravitational potential to obtain extra
conditions. First, from the condition that the gravitational potential should be
regular at the center we obtain
\be
\delta\alpha'=0 {\rm ~at~} r=0, \label{eqa14-2}.
\ee
Second because the gravitational potential should fall off as $1/r$ beyond the stellar
surface, we have
\be
\delta\Phi'+\frac{\delta\Phi}{r} = 0.
\ee
When we apply the above equation at the stellar boundary ($r=r_s$), we get
\be
\delta\alpha' = -\frac{\delta\alpha^2 -1}{2r \delta\alpha} {\rm ~at~} r=r_{s}.\label{eqa14-3}
\ee

Since \refeqns{eqa11-1}--\eqref{eqa11} and \eqref{eqa9} are coupled,
we use an iterative method to solve them.

For the case of the Cowling approximation, which assumes that the gravitational potential is
fixed ($\delta\alpha=0$), the equations simplify considerably:
\ba
&&\delta P = -\Gamma P_{0}[r^{-2}\alpha_{0}^{3}(r^2\alpha_0^{-3}\xi)'] -\xi P_0' \label{eqa12-1}\\
&&\delta \rho = -(\rho_0+P_{0})[r^{-2}\alpha_{0}^{3}(r^2\alpha_0^{-3}\xi)'] -\xi \rho_0'
\label{eqa12-2}\\
&&(\rho_0+P_0)\alpha_0^{-4}\sigma^2\zeta=\delta P' +(\delta\rho+\delta
P)\alpha_0^{-1}\alpha_0'
\label{eqa12}
\ea
If we compare the above equations with \refeqns{eqa11-1}--\eqref{eqa11},
we observe that every coefficient of $\delta\alpha$ is negative: therefore, as mentioned in the
main text,
$\sigma$ increases when we apply the Cowling approximation.

\begin{figure*}
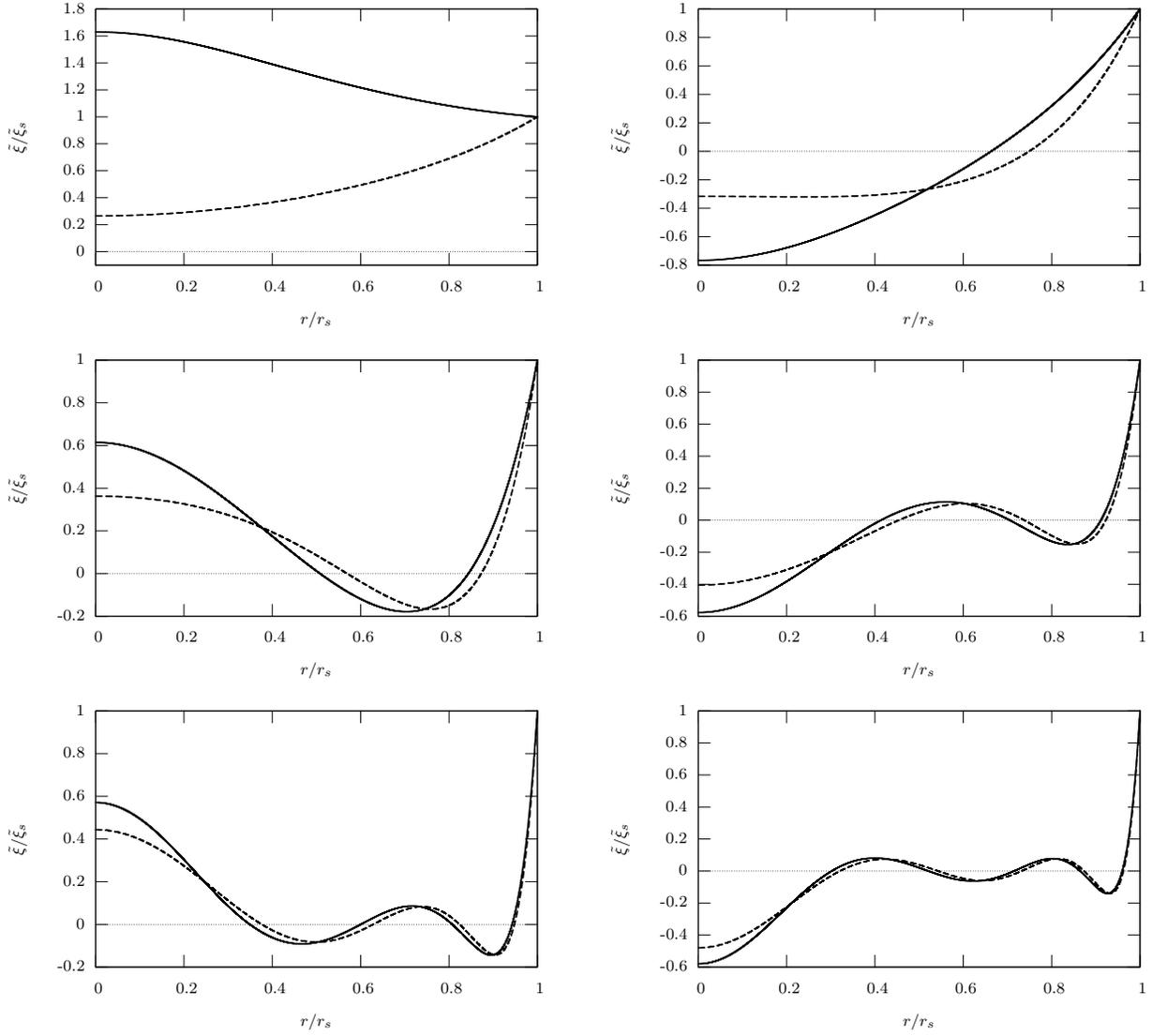

\begin{center}
\begin{tabular}{cc}
\scalebox{0.8}{\input{figure/xi_F.tex}}&
\scalebox{0.8}{\input{figure/xi_H1.tex}}\\
\scalebox{0.8}{\input{figure/xi_H2.tex}}&
\scalebox{0.8}{\input{figure/xi_H3.tex}}\\
\scalebox{0.8}{\input{figure/xi_H4.tex}}&
\scalebox{0.8}{\input{figure/xi_H5.tex}}
\end{tabular}
\caption{Radial pulsation eigenfunction of a 1.4\solarmass star. The equation of
state that we use is the
polytropic one with $K=100$ and $N=1$. In this figure, $\tilde{\xi}=\xi/r$ and
$\tilde{\xi}_s=\tilde{\xi}(r=r_s)$ where $r_s$ is the surface radius. The dashed and
solid lines
represent the results with and without the Cowling approximation, respectively.
Each panel shows different modes (top-left ($F$), top-right ($H_1$), middle-left ($H_2$), middle-
right ($H_3$), bottom-left ($H_4$) and bottom-right ($H_5$)) which have different oscillation
frequencies.}
\label{fig09}
\end{center}
\end{figure*}

We show the solution for $\xi/r$ for the $1.4\solarmass$ star with $K=100$ and $N=1$
with and without the Cowling approximation in Figure \ref{fig09}.
The $\sigma$ values corresponding to each mode are summarized in Table \ref{tb3} which
appears in the main text.

\end{appendix}

\label{lastpage}

\end{document}